\documentclass[prb,twocolumn,showpacs,amsmath,amssymb]{revtex4}% Physical Review B

\usepackage{graphicx}% Include figure files
\usepackage{dcolumn}% Align table columns on decimal point
\usepackage[xdvi]{epsfig}
\usepackage[usenames]{color}
\usepackage{amsfonts}
\usepackage{txfonts}
\usepackage{subfigure}
\usepackage[ansinew]{inputenc}

\definecolor{MyLightMagenta}{rgb}{0.1,0.2,0.8}
\let\oldmarginpar\marginpar
\renewcommand\marginpar[1]{\-\oldmarginpar[\raggedleft\large{\color{MyLightMagenta} \vspace{-12pt} #1}]{\raggedright\large{\color{MyLightMagenta} \vspace{-12pt} #1}
}}
\newcommand{\rem}[1]{}

\begin{document}

%\preprint{APS/123-QED}

\title{First-principles study of the effect of Fe impurities in MgO at geophysically relevant pressures}

\author{Donat J. Adams}
\email{donat.adams@cea.fr}
\affiliation{Dep. Materials Sciences,
Lab. Crystallography, ETH Z\"{u}rich, Switzerland}
\altaffiliation{CEA, DAM, DIF, F 91297 Arpajon, France}

%\author{Artem R. Oganov}
%\affiliation{Department of Geosciences, Department of Physics and Astronomy, and New York
%Center for Computational Sciences, State University of New York, Stony Brook, NY 11794-2100 \\
%and\\
%Geology Department, Moscow State University, 119992 Moscow, Russia}

\author{W.M. Temmerman}
\author{Z. Szotek}
\affiliation{Daresbury Laboratory, Daresbury, Warrington WA4 4AD, United Kingdom}
\date{\today}

\begin{abstract}
The self-interaction corrected local spin density (SIC-LSD)
formalism and the standard GGA treatment of the exchange-correlation
energy have been applied to study the collapse of the magnetic
moment of Fe impurities in MgO. The system Mg$_{1-x}$Fe$_x$O is
believed to be the second most abundant mineral in the Earth's lower
mantle.
%The SIC-LSD is used to study the
%effects of strong electron correlation on the Fe-3\emph{d} orbitals,
%in particular the high spin (HS) to low spin (LS) transition. Fe
%concentrations studied are between 3.125\% and 100\%. At low
%concentration ($x\leq$ 12.\%) the resulting transition pressure
%range between 26~GPa and 63~GPa while at high concentration
%(100\%$\geq x\geq$ 12.\%) they are between 83~GPa and 296~GPa.
We confirm the experimentally found increase of the critical
pressure upon iron concentration. Our calculations using standard
GGA for a fixed Fe concentration show that different arrangements of
Fe atoms can remarkably shift the transition pressure of the high
spin (HS) to low spin (LS) transition. This could explain the
experimentally found broad transition regions. Our results indicate
that the HS-LS transition in Mg$_{1-x}$Fe$_x$O is first order. We
find that SIC-LSD fails to predict the divalent Fe configuration as
the lowest energy configuration and discuss possible reasons for it.
\end{abstract}

%\pacs{Valid PACS appear here}% PACS, the Physics and Astronomy% Classification Scheme.

%\keywords{Suggested keywords}%Use showkeys class option if keyword %display desired

\maketitle

%\lowercase{via} \textbackslash \protect\\ \url{} \onlinecite{feyn54}
\section{Introduction \label{sec:int}}

The high spin (HS) to low spin (LS) transitions in
magnesiow\"{u}stite, Mg$_{1-x}$Fe$_x$O, and w\"{u}stite,
Fe$_{1-x}$O, are of great geophysical importance.
Magnesiow\"{u}stite  is believed to be the second most abundant
mineral in the Earth's mantle.\cite{ono2005, anderson1989} While MgO
was thoroughly studied both in theory and
experiment\cite{oganov2002,oganov2003, duffy1995, jaffe2000,
zhang1991, chang1984, karki2000, drummond2002, kalpana1995,
causa1986, strachan1999, dewaele2000, speziale2001, mehl1988} at
high pressures, few theoretical studies exist for Mg$_{1-x}$Fe$_x$O
mainly because of the difficulty to treat the effects of strong
correlation for the Fe-3\emph{d} orbitals.\cite{persson2006,
tsuchiya2006,fang1999} The magnetic phase transition could strongly
influence the partition coefficient of Fe between MgO and MgSiO$_3$
perovskite and postperovskite,\cite{badro2003} dramatically change
radiative conductivities,\cite{goncharov2006} and lead to a
hardening of the materials.\cite{gaffney1973}
%\\
However, magnesiowüstite Mg$_{1-x}$Fe$_x$O and wüstite Fe$_{1-x}$O
have been studied in numerous experimental
papers.\cite{fei1992,lin2005,badro2003,welberry1997, welberry1995,
carel1998, gavarri1981b} The main problem has been, that wüstite
exists only as a non-stoichiometric compound.\cite{ding2005,
ding2005b,ding2005c, ono2007, fang1999}
%
%\hevo{
There is a general consensus that its rocksalt-type structure
contains a fully occupied O$^{2-}$ sublattice, while Fe exists
predominantly in the form of Fe$^{2+}$, with some Fe$^{3+}$, as well
as vacancies and interstitials, exhibiting a short range order.
%}{1}
The resulting clusters arrange with a long
range order, which -- depending on the density of interstitials --
can be commensurate or incommensurate.\cite{welberry1997,welberry1995,carel1998,gavarri1981b}

For Fe$_{1-x}$O \citet{jacobsen2005} reported a transition from a
cubic to a rhombohedral phase at a pressure between 22.8 GPa and
27.7 GPa at room temperature. At low temperatures, on the other
hand, \citet{struzhkin2001} found a phase transition from a cubic to
a rhombohedral structure with a N\'{e}el-temperature of $T_N=198$~K
at ambient pressure. At ambient temperature this transition is
shifted to a pressure of 15~GPa.

At pressures of 85 to 143~GPa a magnetic high-spin to low-spin
transition was observed in several experiments.\cite{ono2007,
fang1999, pasternak1997} Some studies find a broad transition
region,\cite{pasternak1997} while in others\cite{ono2007} the
transition is first order. Differential stress might have smeared
out the phase transition over a broad region in some studies or
small compositional differences within the sample could have changed
the transition behavior.

Recent developments in high pressure physics allowed to investigate
the LS-HS transition in magnesiow\"{u}stite. \citet{badro2003} and
\citet{lin2005} could prove what was suggested a long time
before:\cite{anderson1989} the complete electronic rearrangement of
Fe in Mg$_{1-x}$Fe$_x$O under pressure, while the NaCl-type
structure remains stable across the phase transition. The transition
pressures are in reasonable agreement (49 to 75~GPa for
$x=0.15$,\cite{badro2003} 50 to 60~GPa for $x=0.25$\cite{lin2005}).
The transition is linked to a hardening of the materials and a
decrease of the molar volume (across the phase transition:
$V_{0\mathrm{LS}}/V_{0\mathrm{HS}} = 0.904$, bulk modulus:
$K_{0T}^{\mathrm{HS}} = 160.7$~GPa, $K_{0T}^{\mathrm{LS}} = 250$~
GPa\cite{lin2005}).

In another study the HS-LS transition in Mg$_{1-x}$Fe$_x$O was
studied for different iron concentrations ($x = 0.2$, 0.5
 and 0.8).\cite{speziale2005} The transition pressure turns out
to depend linearly on the Fe concentration and is 40, 60 and 80 GPa,
respectively. For Fe$_{0.97}$O the transition pressure is shifted to
90~GPa. %\hevo{
At high Fe concentration ($x=17$\%) \citet{badro1999} in an X-ray
emission study find that the HS state is stable up to 143~GPa,
whereas \citet{pasternak1997} conjecture from Mössbauer spectroscopy
a strong temperature dependence of the transition pressure. They
found the transition to be completed only at 120~GPa (Fe$_{0.94}$0)
at 450~K.

In the present paper we study the effect of Fe impurities in MgO at
geophysically relevant pressures, using self-interaction corrected
local spin density (SIC-LSD) method which allows to treat the
localized $d$ electrons of Fe on equal footing with the other
itinerant electrons. In addition, we use GGA approach and the
VASP code\cite{kresse1996} to study the influence of possible Fe
impurity clustering in the MgO supercells on the relevant transition
pressures. The aim is to realize different Fe concentrations in MgO
and investigate their effect on the experimentally observed
properties.

The outline of the paper is as follows. In Section II, we briefly
describe the SIC-LSD methodology. Section \ref{sec:res} concentrates
on the discussion of the present SIC-LSD and GGA results, while
Section \ref{sec:con} concludes the paper.

\section{ Methodology \label{sec:meth} }

\subsection{The self-interaction corrected LSD \label{sec:meth:sic}}

The standard LSD approximation for the exchange-correlation energy
introduces an unphysical interaction of an electron with itself, the
so-called self-interaction (SI). It is the aim of the self-interaction
corrected local spin density approximation to construct a self-interaction
free energy functional

\begin{equation}\label{eq:sic:1}
E_{\mathrm{SIC}}= E_{\mathrm{LSD}}- \sum_\alpha
\delta_\alpha^{\mathrm{SIC}} ,
\end{equation}

\noindent
where $\alpha$ numbers the occupied orbitals and the
self-interaction correction (SIC), $\delta_\alpha^{\mathrm{SIC}}$, of
the orbital $\alpha$ is

\begin{equation} \label{eq:sic:2}
\delta_\alpha^{\mathrm{SIC}} = U[n_\alpha]+
E_{\mathrm{xc}}^{\mathrm{LSD}}[\bar{n}_\alpha]\;.
\end{equation}

%\noindent
It is known that the exact exchange-correlation energy
$E_{\mathrm{xc}}[\bar{n}_\alpha]$, depending on the spin density
$\bar{n}_\alpha$, for the case of a single electron orbital with
electron density $n_\alpha$, cancels the Hartree energy $U[n_\alpha]$
identically:

\begin{equation} \label{eq:sic:3}
U[n_\alpha]+E_{\mathrm{xc}}[\bar{n}_\alpha]=0\;.
\end{equation}
In the case of approximate exchange-correlation energy functionals
this cancelation can be guaranteed by Eq. \ref{eq:sic:1}.

Varying the above
SIC-LSD energy functional with respect to the orbital spin densities,
with the constraint that the $\phi_{\alpha}$'s form a set of orthonormal
functions, one gets the SIC-LSD generalized eigenvalue equations
%
% SIC equations  eq-sic
%
\begin{eqnarray}
\label{eq:eq-sic}
%
% H_{\alpha} \mid \phi_{\alpha} > &=& \left( - \nabla^{2}
% + V_{{\rm eff},\alpha \sigma}^{\rm SIC-LSD}({\bf r}) \right) \mid \phi_{\alpha}> \nonumber\\
H_{\alpha} \mid \phi_{\alpha} > & =& \left( H_{0\sigma} +V_{\alpha}^{SIC}({\bf r}) \right) \mid \phi_{\alpha}> = \sum_{\alpha'}
 \lambda_{\alpha \alpha'} \mid \phi_{\alpha'}> ,
 \end{eqnarray}
with $H_{0\sigma}$ being the orbital independent LSD Hamiltonian. The
Lagrangian multipliers $\lambda_{\alpha \alpha'}$ are used to secure the fulfilment of
the orthonormality constraint.

Due to the orbital dependent SIC potential, $V_{\alpha}^{SIC}$, the SIC energy
functional is not stationary with respect to infinitesimal unitary transformations among the orbitals.
% of the basis set.
The so-called localization criterion
 %
 % localization criterion
 %
\begin{equation}
<\phi_{\beta} \mid V_{\alpha}^{SIC} - V_{\beta}^{SIC} \mid
\phi_{\alpha}> = 0 \quad \forall (\alpha,\beta)
\end{equation}
has to be fulfilled to ensure that the solutions of the SIC-LSD
equations (\ref{eq:eq-sic}) are most optimally localized to reach the absolute
minimum of the SIC-LSD functional (\ref{eq:sic:1}).

The SIC-LSD approach is fully \emph{ab initio} and introduces no adjustable
parameters, either for the delocalized (band-like) or localized
electrons. For extended states the SIC vanishes. The SIC-LSD
formalism has the advantage that it allows to compare different
valence states and spin configurations of the same atom.
The nominal valence, $N_{val}$, in the SIC-LSD approach is
defined as
%
% valency eq:val
%
\begin{equation} \label{eq:val}
N_{val} = Z -N_{core} - N_{SIC}  \, ,
\end{equation}
where $Z$ is the atomic number, $N_{core}$ is the number of core (and
semicore) states and $N_{SIC}$ is the number of self-interaction
corrected states. The ground state valence is the one defined by the
ground state energy.
In our application of SIC-LSD to (Mg,Fe)O all possible electron
configurations of the Fe atom with five and six SI-corrected
$d$ orbitals have been considered.\footnote{The nominal valence
$N_{\mathrm{val}}=Z-N_{\mathrm{core}}-N_\mathrm{SIC}$ does not
always correspond to the chemical valence. For example, in the case of
the conventional LSD scheme the valence of Fe would be
$N_{\mathrm{val}}=Z-N_{\mathrm{core}}-N_\mathrm{SIC}=26-18-0=8$.}
%They were labeled a--n for Fe$^{2+}$ and A--N for Fe$^{3+}$ and
The most energetically relevant ones are given in
Table~\ref{tab:spinc}. The energy functional without any SIC states
is simply equivalent to the standard LSD approximation and labelled
lsd in the following. Thus LSD is a local minimum of the SIC-LSD
functional, corresponding to the case when all the electrons are
treated as itinerant and described by the Bloch wave functions. The
choice of the SIC orbital configuration, i. e. the number and
symmetry of localized vs. delocalized states, can significantly
influence the corresponding total energy. The electronic
configuration giving rise to the most negative total energy,
with the most optimally localized orbitals, defines the absolute energy
minimum of the SIC-LSD energy functional as well as the corresponding
valency.

\begin{table}
\caption{\label{tab:spinc} The most relevant spin configurations of
Fe, labeled HS and LS according to Ref.~\onlinecite{speziale2005}.
Here HS state corresponds to the case where SIC is applied to all
the majority and one minority electrons. For realizing LS state,
SIC is applied to the three majority and three minority t$_{2g}$
electron states.
The HS$^{3}$ stands for the trivalent Fe configuration, where all
the majority $d$ electrons are treated as localized, by applying
SIC.  The calculations with applying no SI-corrections have been labeled
lsd. All possible electronic configurations with five or six
SI-corrected orbitals namely trivalent and divalent ions have been
considered in the present study. However, the ones listed below
correspond to the lowest enthalpies. Here 1 means that a given state
has been SI-corrected, and 0 otherwise.}
\begin{ruledtabular}
\begin{tabular}{c|c|c|c|c|c||c|c|c|c|c|c}

                  & \multicolumn{5}{c||}{Majority channel} & \multicolumn{5}{c|}{Minority channel}\\
                      \cline{2-11}
    &  t$_{2_g}\!\!$  & t$_{2_g}\!\!$ & e$_g\!\!$  & t$_{2_g}\!\!$ & e$_g\!\!$   &  t$_{2_g}\!\!$  & t$_{2_g}\!\!$ & e$_g\!\!$  & t$_{2_g}\!\!$ & e$_g\!\!$ \\
    \hline
HS & 1 & 1 & 1 & 1 & 1   & 1 & 0 & 0 &  0 & 0 \\               \hline  %b
LS & 1 & 1 & 0 & 1 & 0   & 1 & 1 & 0 &  1 & 0 \\              \hline   %n
HS$^3$ & 1 & 1 & 1 & 1 & 1   & 0 & 0 & 0 &  0 & 0 \\

\end{tabular}
\end{ruledtabular}
\end{table}

The SIC-LSD approach used in this work has been implemented in the
linear muffin-tin orbitals (LMTO) band structure method with the
atomic sphere approximation (ASA).\cite{temmerman1998, andersen1984,
andersen1975} The slightly overlapping ASA spheres approximate the
polyhedral Wigner-Seitz cell and the sum of their volumes equals to
the volume of the actual unit cell. The ASA does not allow for
atomic relaxations, as different ionic arrangments will lead to
different sphere overlaps. In the LMTO-ASA approach empty spheres
(E) can be introduced in order to increase the space filling and to
minimize the overlap.

In all the calculations performed with the LMTO-ASA method we have
treated the valence states of the Mg-3\emph{s}, Mg-3\emph{p},
%Mg-3\emph{d},
O-2\emph{s}, O-2\emph{p}, Fe-3\emph{d}, Fe-4\emph{s},
Fe-4\emph{p}, E-1\emph{s} as the so-called low waves, while the
Mg-3\emph{d}, O-3\emph{d} and E-2\emph{p} states have been represented as
intermediate waves.\cite{lambrecht1986} All the electrons in lower
shells have been put in the core but allowed to relax.
%Among the valence states low, intermediate and high
%waves are distinguished in order to reduce the size of the
%Hamiltonian. High waves are considered only implicitly.
The separation of valence electrons into the low and intermediate
waves has been done for the purpose of reducing the size of the
eigenvalue problem. The secular equation that is fully diagonalized
is constituted by the low waves and provides the eigenvalues and
eigenvectors. The intermediate waves enter the Hamiltonian through
their tails, partly retaining their true characteristics.

For both magnesiow\"{u}stite and w\"{u}stite, we have assumed the
ideal rock salt structure. The ASA radii of MgO were adjusted in
order to reproduce the experimental band gaps of
7.83~eV.\cite{whited1969}
%and 8.9~eV \citep[HF calculation]{pantelides1974} for the lattice constant (4.211~\AA) calculated by \citet{oganov2003}.
Best agreement was found with radii of 1.34646~\AA, 1.05313~\AA,
0.751847~\AA~ for Mg, O and the interstitial spheres E, with the
resulting LDA band gap of 4.7~eV. The total volume of the spheres
was equal to the volume of the cell for all the SIC calculations
while the ratios of the ASA radii were kept constant. The ASA radius
of Fe was equal to the one of Mg.

To realize different concentrations of Fe impurities in the systems
studied, three supercells have been used, namely consisting of four,
eight and 32 formula units of MgO. By substituting up to four Mg
atoms by Fe atoms one could realize Fe concentrations ($x= \frac{
n_{\mathrm{Fe}} }{ n_{ \mathrm{ cations }}}$) of 3.125\%, 12.5\%,
25\%, 50\% and 100\%.
%
%%%%%%%%%%%%%%%%%%%%%%%%%%%%%%%
%For any of these Fe concentrations the smallest possible supercell of the ideal NaCl-type structure has been used.
%%%%%%%%%%%%%%%%%%%%%%%%%%%%%%%
For Fe concentrations of 3.125\%, 12.5\%, 25\%, we have chosen the
smallest possible supercell of the ideal FCC/NaCl, namely replacing
one Mg atom by one Fe atom respectively in the supercells consisting
of 32, eight and four formula units. The concentrations of 50\% and
100\% have been realized using a supercell of eight atoms in total
and replacing respectively two and four Mg atoms by Fe atoms.
Consequently, both ferromagnetic (FM) and antiferromagnetic (AFM)
orders as well as charge disproportionation could be studied. This
corresponds to an Fe arrangement with maximal distances between the
Fe atoms and would provide the lowest transition pressures.
Regarding the AFM structure, the chosen supercells allowed to
realize only the so-called AF1 order where the magnetic spins are
arranged in parallel in the (001) planes, but are anti-aligned
between the planes.
%\hevo{This corresponds to a Fe arrangement with
%maximal distances between the Fe atoms and would provide the lowest
%transitions pressures.}{1}
The number of k-points has been chosen inversely proportional to the
size of the supercell (eight, 16, 64 atoms) and was
$16\times16\times16$, $8\times8\times8$ and $4\times4\times4$,
respectively. This gives energy differences precise to within
1.7$\times$10$^{-6}$~eV/atom (1.25$\times$10$^{-7}$~Ryd/atom).

\subsection{Pressure determination}
%\subsection{Pressure determination in the LMTO-ASA\label{sec:meth:pres}}

The pressures and enthalpies, of relevance to the present study, can
be determined from the LMTO-ASA total energy, E, vs. volume, V,
calculations through invoking the equation of state (EOS) and
fitting to a number of calculated data points.\footnote{We have used
a standard procedure such as provided with the EXCITING code at
http://exciting.sourceforge.net.} Specifically, the third-order
Birch Murnaghan EOS has been used,\cite{poirier2000, birch1947}
namely

\begin{eqnarray} \label{eq:pres:1a}
E(x)& = E_0&+ \frac{9 B_0 \, V_0}{16} \Big (B'(x^{2/3}-1)^3 \notag \\
& &+(x^{2/3}-1)^2\times (6-4 x^{2/3}) \Big ),
\\ \label{eq:pres:1b}
P(x)&= &\frac{3 B_0}{2} \left (x^{7/3}-x^{5/3} \right) \notag  \\
& & \times \left (1+\frac{3}{4}\,(B'-4)\times (x^{2/3}-1 ) \right ),
\\
x& = \frac{V_0}{V} \;  ,
\end{eqnarray}

\noindent allowing to compute the enthalpy $H$ at any volume as
$H(V)=E(V)-P(V)\,V$, with $P$ denoting the pressure. In the above,
$V_{0}$ stands for the equilibrium volume, $B_0$ for the bulk
modulus and $B'$ for its first pressure derivative.

The calculation of enthalpy differences, $\Delta H(P)$, can be
accomplished through the numerical inversion of the pressure in Eq.
\ref{eq:pres:1b}. This allows to compare enthalpies of two spin
configurations labelled $s$ and $s'$ at a given pressure:

\begin{equation} \label{eq:pres:2}
\Delta H^{(s,s')}=H^s(V^s)-H^{s'}(V^{s'}) \; .
\end{equation}

To determine the theoretical equilibrium volumes and total energies
for all the studied scenarios, the lattice parameters have been
sampled in the range from -~10\% up to +~16\% (relative to the
zero-pressure value) in steps of 2\%. The resulting volume increase
or decrease of -33.1\% to 40.7\% typically corresponds to a pressure
range of -20 GPa to 285 GPa.\footnote{The enthalpy differences have
been fitted to the second order polynomials. However, the error
resulting from these fits and the numerical inversion of
Eq.~\ref{eq:pres:1b} have given rise to total errors of the order
10$^{-8}$ -- 10$^{-12}$~eV/atom.}
%Note that for the comparison of enthalpies no
%additional full potential calculation is needed.

\section{Results and Discussion} \label{sec:res}

\subsection{(Mg,Fe)O in the SIC-LSD framework \label{sec:res:sic}}

%The introduction of supercells containing of 4, 8 and 32 formula units of MgO and a subsequent substitution of up to 4 Mg ions by Fe allows to study Fe concentrations of $x= \frac{ n_{\mathrm{Fe}} }{ n_{ \mathrm{ cations }}} = 3.125$~\%, 12.5~\%, 25\%, 50\% and 100\%. For any concentration the smallest possible supercell of the ideal FCC NaCl was chosen (Fig. \ref{pic:res:dh_sic_mgo64})\\

In the application of SIC-LSD to Fe-doped MgO, we have studied 29
different localization-delocalization scenarios for three different
supercells, various Fe concentrations, and symmetries of localized
Fe $d$ states. Among all the cases we have identified four
energetically relevant configurations, namely the three SIC
configurations listed in Table \ref{tab:spinc} as well as the LSD
solution. This finding has been independent of the actual Fe
concentration. The HS configuration is the high-spin divalent
scenario,
%\hevo{where
%all majority Fe $d$ states and one minority Fe $t_{2g}$ state}{3}
where all majority Fe $d$ states and one minority Fe $t_{2g}$ state
are treated as localized through invoking SIC, giving rise to an
insulating state. The HS$^3$,
trivalent, configuration corresponds to the scenario where only the
majority Fe $d$ states are corrected for the self-interaction, and
thus localized, constituting a half-metallic state. The LS state
corresponds to the case where both
majority and minority Fe $t_{2g}$ states are localized by SIC. This
naturally leads to a non-magnetic and insulating configuration.
%It is a common feature of all sampled cells corresponding to different Fe-concentrations,
%that the HS, HS$^3$, LS and lsd belong to the energetically lowest lying structures.

\begin{figure}
\begin{center}
\subfigure[]{ %Real and imaginary.
\includegraphics[width=0.45\textwidth]{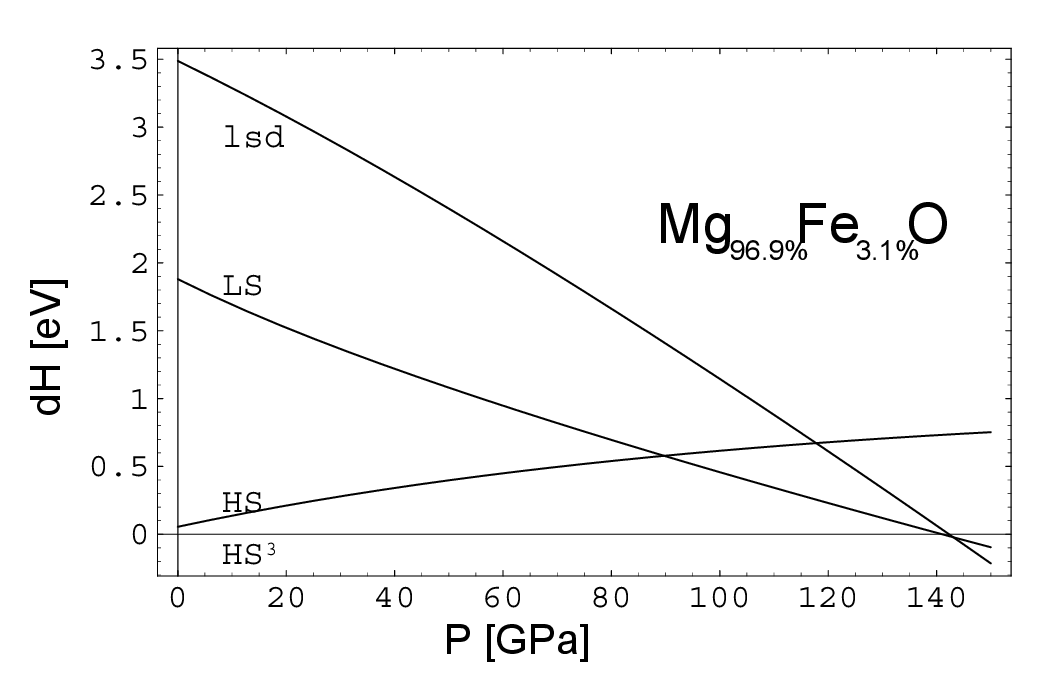}   \label{pic:res:dh_sic_mgo64}}
\subfigure[]{ %Amplitude and phase.
\includegraphics[width=0.45\textwidth]{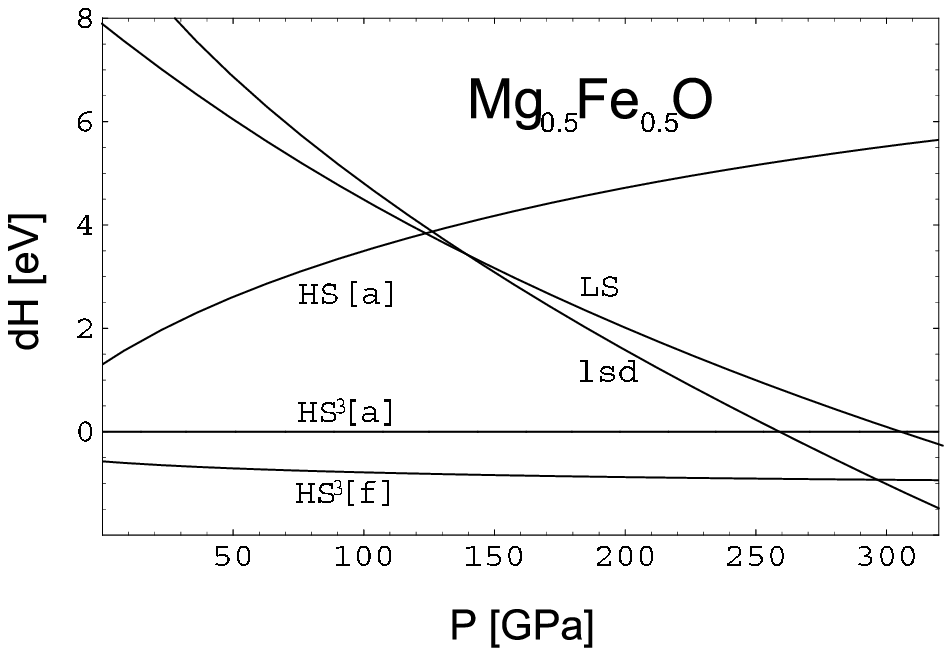}  \label{pic:res:dH_mgo8-fe2}}
%\includegraphics[width=0.45\textwidth]{graph/dH_mgo8-fe2}  \label{pic:res:dH_mgo8-fe2}}
%\subfigure[]{ %Amplitude and phase.
%\includegraphics[width=0.45\textwidth]{graph/dh_feo8}   \label{pic:res:dh_feo8}}
\caption{\label{pic:res:dhs} Enthalpy differences at different Fe
concentration $x$. \subref{pic:res:dh_sic_mgo64} $x=$~3.125\%. The
HS$^3$ configuration appears to be energetically most favorable up
to a pressure of 145~GPa, where a phase transition is predicted to
the state where all the Fe-3\emph{d} delocalize (lsd). The commonly
assumed HS configuration is less energetically favourable, and at
the pressure of 90~GPa the electrons completely rearrange to the
non-polarized configuration LS, although the localization of the
electrons is not lost. \subref{pic:res:dH_mgo8-fe2} $x=$~50\%. Two
Fe atoms were introduced into a cell containing 8 atoms in total
(two Mg, two Fe, four O). Parallel and antiparallel alignments of
the spin moments is denoted in brackets [f] and [a], respectively.
More than 10 different arrangements with asymmetric charge
configurations on the Fe atoms were also tried out, but resulted in
high enthalpies. They are omitted in the plot for more clear
readability.
  }
\end{center}
\end{figure}

From the enthalpy calculations we find, in disagreement with
experimental evidence, that the HS$^3$ and LSD solutions are
energetically most favorable, respectively at the low and high
pressure regions. This can be seen in Fig.
\ref{pic:res:dh_sic_mgo64}, where the enthalpy differences, with
respect to the HS$^3$ configuration, are plotted as a function of
pressure for all the relevant scenarios and Fe concentration of
3.125\%. The trivalent HS$^{3}$ state constitutes the ground state
solution for all the pressures up to about 140 GPa. However, at very
low pressures the divalent HS state lies very close in energy to the
HS$^{3}$ state and the enthalpy difference, at 0~GPa and 3.125\% Fe
concentration, is $H_{\mathrm{HS}} - H_{\mathrm{HS^3}}=$0.05~eV.
Note that at this Fe concentration the total energies of the HS and
HS$^{3}$ configurations, evaluated at their respective theoretical
volumes, are fully degenerate.

The nominal valence
of the HS$^3$ state, being 3+, is at odds with the experimental
findings, where Fe appears clearly as Fe$^{2+}$ in (Mg,Fe)O (see
e.g. Ref.~\onlinecite{badro2003}) or with a small ratio of
additional Fe$^{3+}$ in Fe$_x$O (see e.g.
Ref.~\onlinecite{ono2007}). Note, however, that the nominal valence
defined in SIC-LSD not always corresponds to the chemical valence.
Although the nominal valences of the trivalent and divalent states
differ by one (namely 3+ vs. 2+), in terms of a simple charge
counting or charge disproportionation, the two ions, corresponding
respectively to the HS$^{3}$ and HS configurations, differ only by
up to 0.1 electron. Similar observation was made by \citet{szotek2003}
for magnetite.
Despite the latter, it is still surprising to find the trivalent,
instead of the divalent, Fe-state as the lowest energy
configuration, in particular for Fe$_{x}$O (with x=1). This is in
variance to the SIC-LSD results obtained for all the other
transition metal monoxides, namely MnO, CoO and
NiO.\cite{kasinathan2006, dane2009}
It is possible that the failure of SIC-LSD to find the divalent
ground state in FeO has its origin in the fact that LSD
substantially overestimates the exchange splitting for the systems
in question. Consequently, the energy gained on localizing an
additional electron to create a divalent Fe ion is not
sufficient to overcome that exchange splitting.
However, since FeO does not
occur in a stoichiometric form, but as Fe$_{1-x}$O, and experiments
indicate existence of both divalent and trivalent ions, one would
need to perform more realistic calculations to establish the
importance of the off-stoichiometry and valence fluctuations for
obtaining the divalent ground state in FeO.
%
%
%In Fig. \ref{pic:res:dh_sic_mgo64} the enthalpy differences, with respect to the HS$^3$ configuration,
%are plotted as a function of pressure, for all the relevant scenarios and Fe concentration of 3.125\%.
%
%It turns out that at low pressures and any
%Fe concentration the commonly assumed\cite{speziale2005} HS (see
%Table~\ref{tab:spinc}), divalent, state competes with the close
%in energy HS$^{3}$, trivalent, configuration.
%They differ by one in the number of d-electrons to which SIC is applied.
%
%Obviously, the HS$^3$ ground state is totally unexpected and in variance to the experimental evidence, and
%this is true for both ends of the
%Mg$_{1-x}$Fe$_{x}$O system. In particular, finding for FeO (x=1) the trivalent Fe configuration as the lower energy solution,
%than the divalent configuration, is in variance to the SIC-LSD results obtained for the other transition
%metal monoxides, namely MnO, CoO and NiO.\cite{kasinathan2006, dane2009}
%However, since FeO does not occur in a stoichiometric form, but as Fe$_{1-x}$O, and experiments indicate existence of both
%divalent and trivalent ions, one would need to perform more realistic calculations to establish the importance of the off-stoichiometry
%for the divalent ground state.

At high pressures, it is the experimentally
indicated\cite{speziale2005} LS configuration that is relevant, with
its nominal valence of 2+ in agreement with the experimentally found
Fe$^{2+}$ in wüstite and magnesiowüstite. The LS configuration
competes with the LSD description of the Fe-3\emph{d} states, as
seen through the corresponding enthalpy difference which at 0~GPa
and 3.125\% Fe concentration is
$H_{\mathrm{lsd}}-H_{\mathrm{LS}}=$0.7~eV (see
Fig.~\ref{pic:res:dhs}), in favour of the LS state, while the LSD
becomes energetically more favorable for pressures in the excess of
140 GPa. At such high pressures the gain in band formation energy on
delocalization of all the Fe $d$ electrons clearly wins with the
localization energy of those electrons. Since the magnitude of SIC
is strongly dependent on the orbitals, one needs to be guided by
energetics in defining the ground state energy and configuration.

%\subsection{Charge disproportiation in (Mg,Fe)O\label{sec:res:sic_charge}}

Introducing several Fe impurities into a supercell
(Mg$_2$Fe$_2$O$_4$ and Fe$_4$O$_4$) opens a possibility of realizing
both parallel (labelled [f], ferromagnetic) and antiparallel
(labelled [a], antiferromagnetic$\equiv$AF1) arrangements of the
spin magnetic moments on those Fe atoms. In Fig.
\ref{pic:res:dH_mgo8-fe2}, the enthalpy differences are plotted for
the above mentioned configurations, and both parallel
and antiparallel arrangements of spins on the Fe atoms in the
relevant supercell with 50\% Fe concentration. For the HS
configuration the [f] and [a] arrangments return the same energies
while for HS$^3$ the [f] arrangement appears always lower in energy
($\Delta E=$0.5~eV for Mg$_{0.5}$Fe$_{0.5}$O and $\Delta E=$1.5~eV
FeO\footnote{$\Delta E$ is given for one supercell containing 2 Mg,
2 Fe and 4 O in the first case and 4 Fe and 4 O in the second.})
than the antiparallel alignment.

%\subsection{HS-LS transition pressures in the SIC-LSD framework \label{sec:res:sic_pres}}

\begin{figure}
\begin{center}
\includegraphics[width=0.45\textwidth]{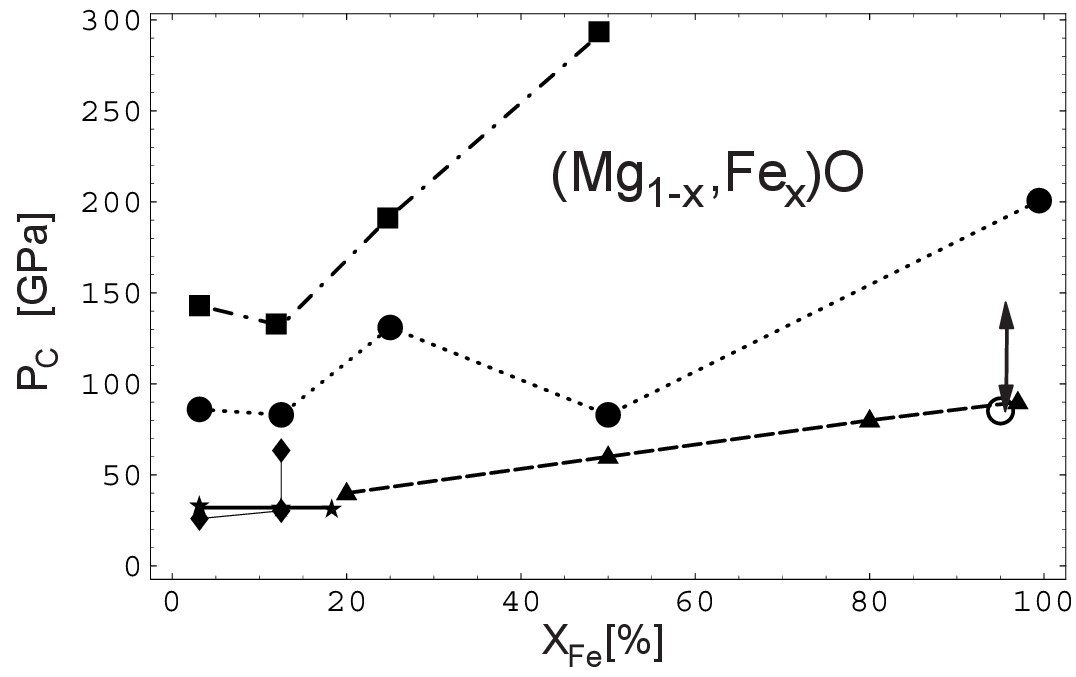}
\caption{\label{pic:res:transpe}
Mg$_{1-x}$Fe$_x$O: Experimental and theoretical transition pressures.
Squares ($\blacksquare$) - SIC-LSD calculations, scenario HS$^3$-lsd (electronic
configuration HS$^3$ at low pressures, no SI-corrections at high pressure);
bullets ($\medbullet$) - SIC-LSD  calculations scenario HS-LS (electronic configuration HS at
low pressures, configuration LS at high pressure); lozenges ($\blacklozenge$) - GGA-PAW
calculations (low concentrations); stars ($\star$) - LDA+U calculations~[\onlinecite{tsuchiya2006b}];
triangles ($\blacktriangle$) - M\"{o}ssbauer spectroscopy,~[\onlinecite{speziale2005}]; arrow
($\updownarrow$) - M\"{o}ssbauer spectroscopy~[\onlinecite{pasternak1997}]; circle ($\medcirc$) -
X-ray diffraction~[\onlinecite{ono2007}].}
\end{center}
\end{figure}

Finally, we consider the dependence of the HS-LS transition pressure on
the concentration of the Fe impurities in a supercell. As can be
seen in Fig. \ref{pic:res:transpe}, our study predicts an increase
of this transition pressure with rising Fe concentration.
%The increase of the HS-LS transition pressure with increasing Fe concentration $x$ is clearly
%seen in our calculations (see Fig.~\ref{pic:res:transpe}).
However, the absolute transition pressures for isostructural phase
transitions, as calculated in this SIC-LSD study, are higher than
the experimental values.\cite{svane2001} Two possible scenarios
can operate at any Fe concentration: a possible transition from
HS$^3$ to lsd or from HS to LS. The second one occurs at
considerably lower pressures than the first one, and %would be the
is the relevant one %if the HS$^3$ configuration was unfavorable (which is
%not the case in the present calculations)
for comparison with experiments, although leading to a
phase diagram in rather poor agreement with experiment. The latter
implies the existence of Fe$^{2+}$, and a low pressure structure
with one minority Fe t$_{2g}$ state and five majority $d$
states.\cite{speziale2005} One might envisage that the HS state
 could possibly become favourable if in the SIC-LSD calculations
the ionic relaxations had been applied. It is obvious that ASA error
can substantially affect total energies, although one would like to hope
that these errors would be less severe when considering only the energy
differences between different configurations.
Also, across the phase transition Fe is
known to reduce its ionic radius and therefore the ionic relaxations
could significantly reduce the enthalpies of the LS structures which
in turn could further decrease the transition pressure.\footnote{In
our calculations the Fe-ASA radii were fixed to those of Mg at
ambient conditions. According to \citet{anderson1989} they are
comparable: $r_{\mathrm{Mg}}=$ 0.72~\AA, $r_{\mathrm{Fe,HS}}=$
0.77~\AA. However, in the LS phase the radius of Fe dramatically
reduces to 0.61~\AA. Due to the LMTO-ASA approach, the reduction of
the ionic radii could not be taken into account.}
%anderson1989, p.338, table 16.1

\subsection{Density of states from SIC-LSD \label{sec:res:dos}}

\begin{figure}[] %Plazierung
%\centering
\subfigure[]{
\epsfig{file=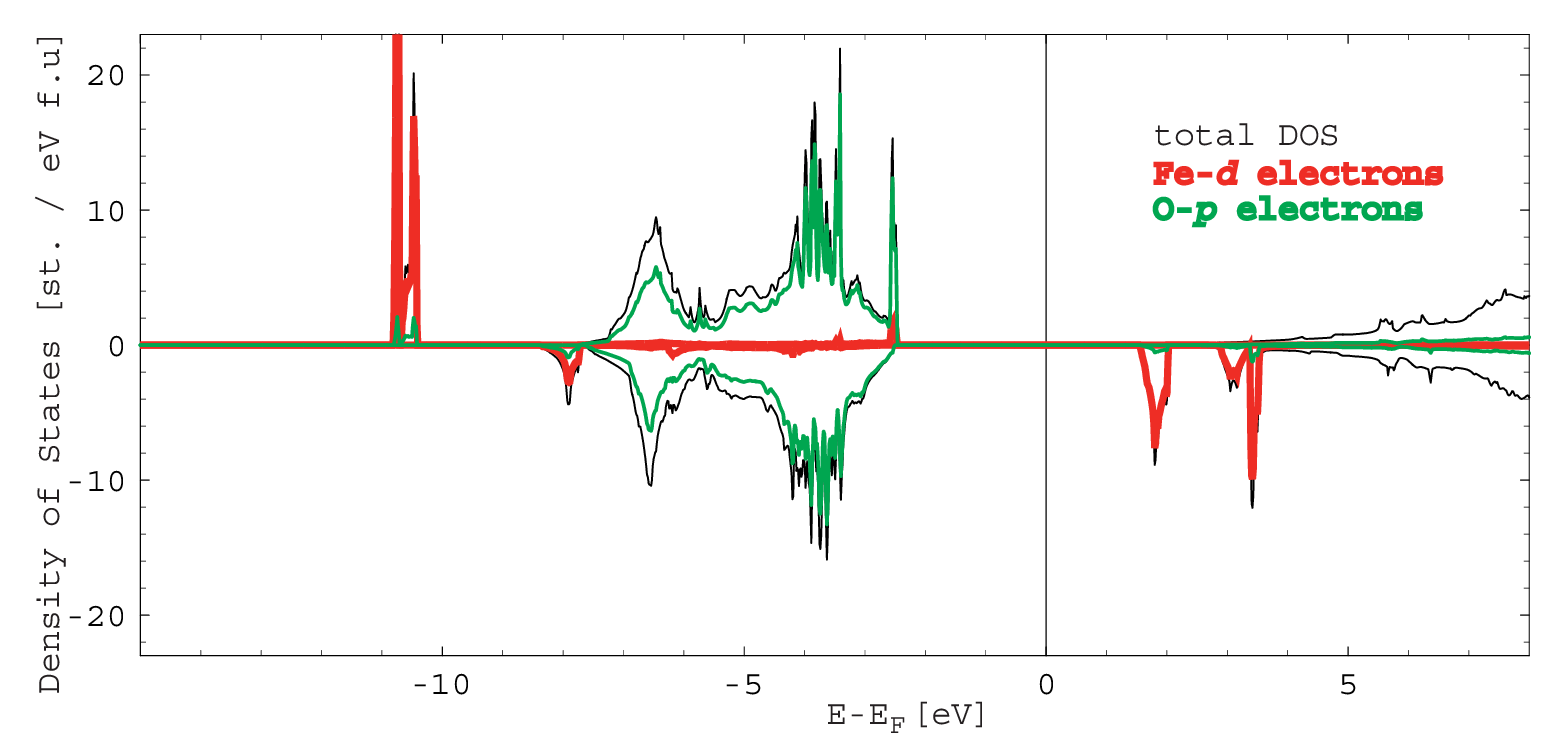, width=0.4\textwidth } \label{pic:res:DOSa}}
\subfigure[]{
\epsfig{file=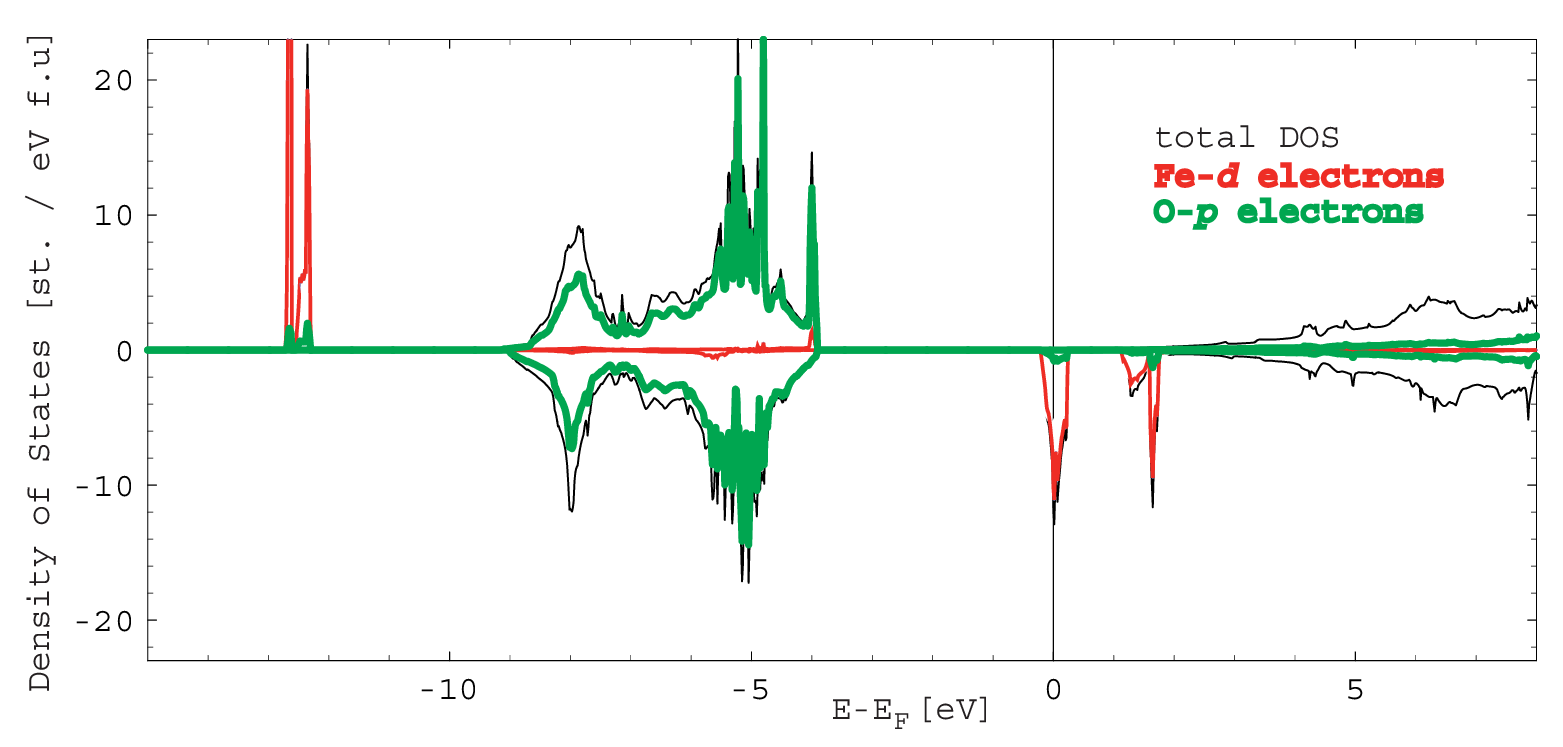, width=0.4\textwidth } \label{pic:res:DOSb}}
\subfigure[]{
\epsfig{file=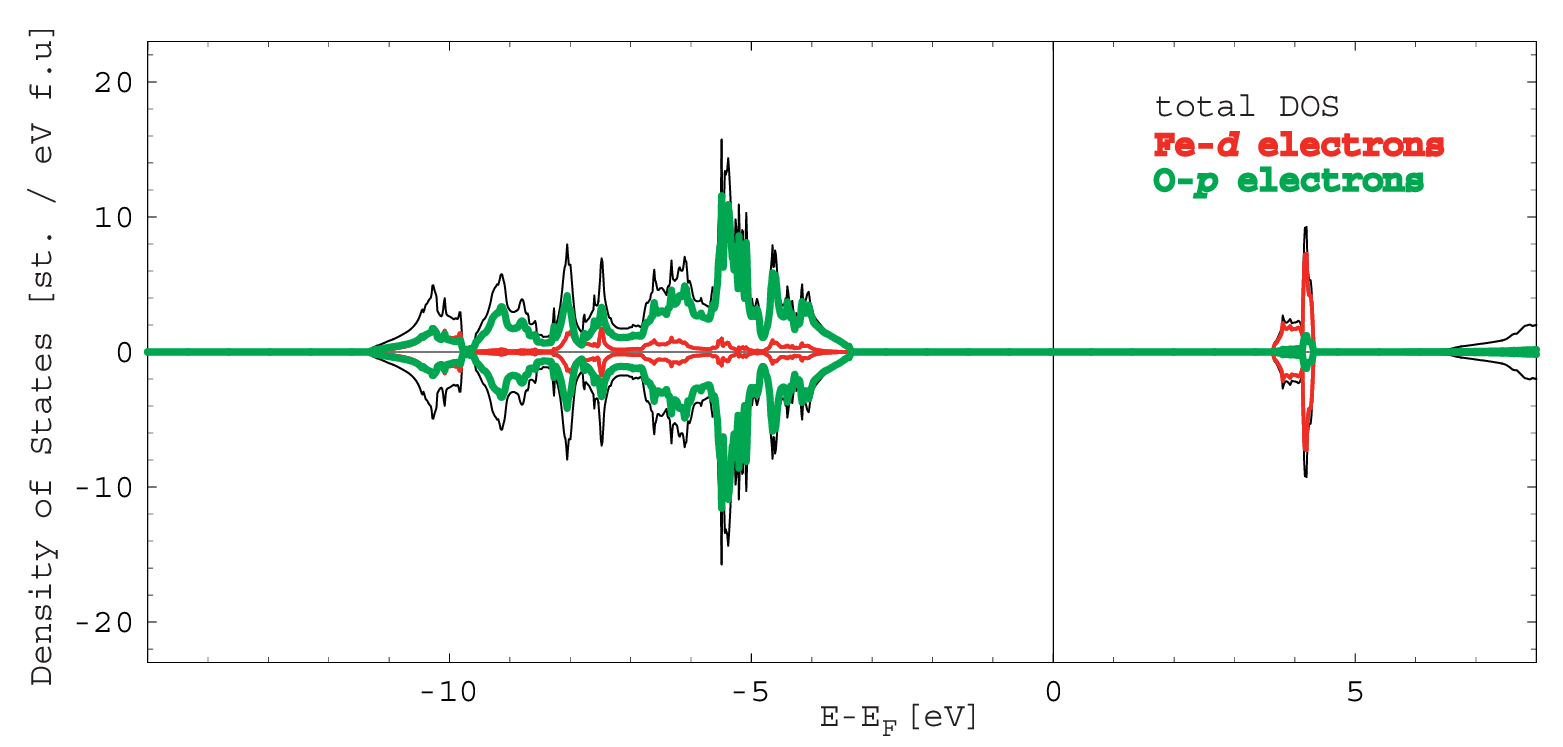, width=0.4\textwidth } \label{pic:res:DOSc}}
\subfigure[]{
\epsfig{file=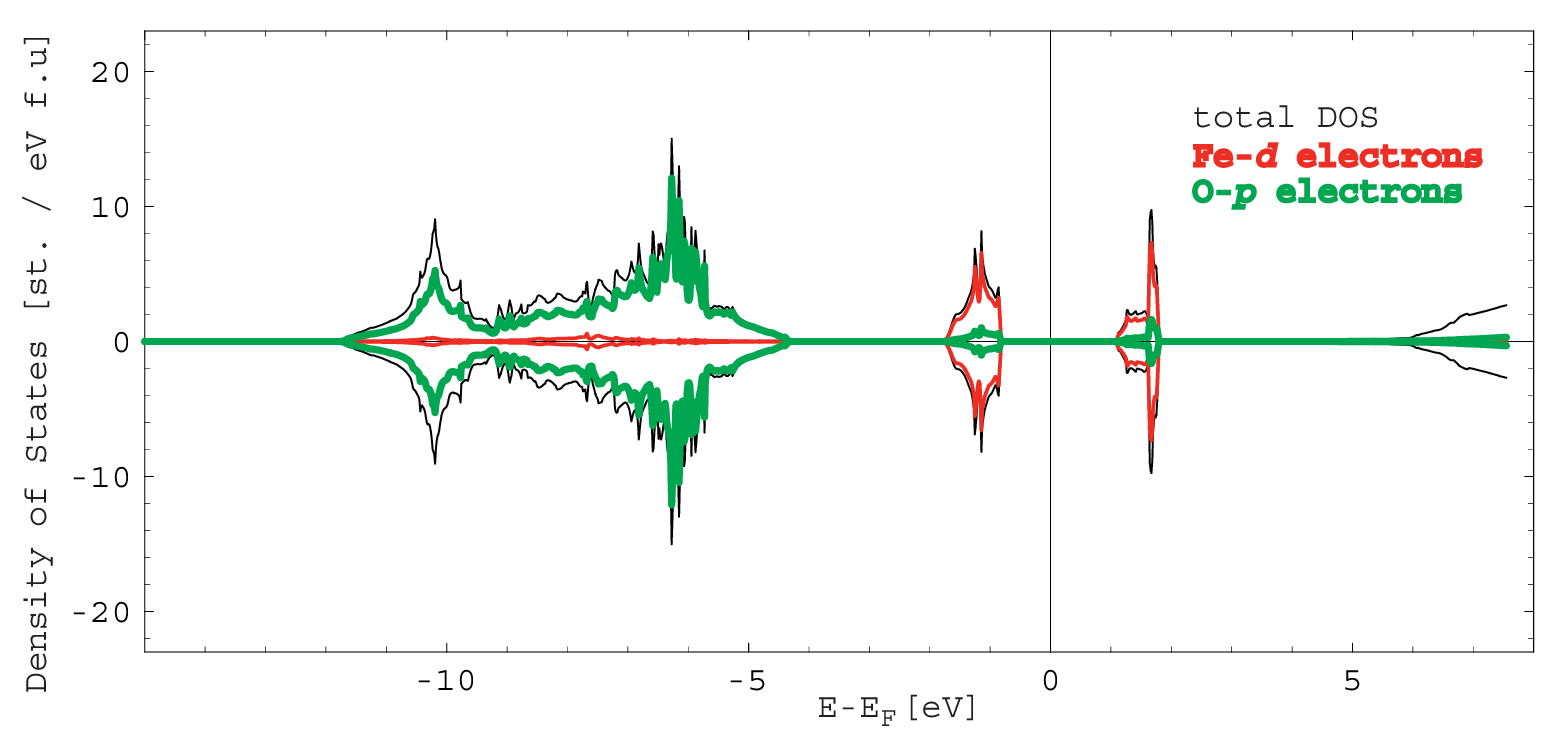, width=0.4\textwidth } \label{pic:res:DOSd}}

\caption{\label{pic:res:DOS} \subref{pic:res:DOSa} Spin-resolved
density of states (DOS) for HS state as calculated from SIC-LSD at a
pressure of 2~GPa and Fe concentration of 12.5\%.
\subref{pic:res:DOSb} Spin-resolved density of states for the
HS$^{3}$ state as calculated from SIC-LSD at a pressure of 2~GPa and
Fe concentration of 12.5\%. \subref{pic:res:DOSc} Spin-resolved
density of states for the LS state calculated at a pressure of
124~GPa and Fe concentration of 12.5\%. \subref{pic:res:DOSd}
Spin-resolved density of states calculated with the standard LSD
approximation at 102~GPa and Fe concentration of 12.5\%. In all the
cases the Fe DOS states is in red, while the total DOS is in
black and the O 2$p$ DOS in green.}
\end{figure}

In Fig. \ref{pic:res:DOS}, we present the densities of states (DOS)
calculated for all the energetically relevant scenarios and the Fe
impurity concentration of 12.5\%.
%The computed densities of states (DOS) show dependence on pressure, Fe concentration and
%the SIC configuration.
As can be seen in the figure, at this intermediate concentration  %(12.5~\%)
the localized Fe-3\emph{d} states form narrow bands. At higher
concentrations of Fe impurities these bands
%are smeared broader
extend over a considerable energy range, e.g. 3-5~eV at 25\% of Fe.
For the HS and HS$^{3}$ configurations, the occupied Fe 3$d$ band
states lie below the predominantly O 2$p$ valence band. As mentioned
earlier, the SI corrections for the HS and HS$^{3}$ configurations
differ by one localized electron more in the former, which is
reflected in the presented DOS (see Figs. \ref{pic:res:DOSa} and
\ref{pic:res:DOSb}).
% :the DOS of the two configurations are essentially the same.
The difference is that the first isolated, minority Fe t$_{2g}$
band, lying just below the valence band, is moved above the valence
band, coinciding with the Fermi level (see Fig.~\ref{pic:res:DOSb}).
While for the HS configuration SIC-LSD delivers, in agreement with
with experiment and other theoretical
considerations,\cite{parmigiani1999, sawatzky1984, zaanen1985} an
insulator of charge transfer character, the HS$^{3}$ configuration
gives rise to a half-metal, with a large band gap in the majority
band, and a metallic behavior in the spin-down channel, with a
partially occupied minority Fe $d$ band at the Fermi level.

At high pressures the LS and LSD configurations are of relevance.
The DOS of the LS is shown in Fig.~\ref{pic:res:DOSc}. It is
characterized by hybridized Fe-3\emph{d} electrons with the
predominantly O 2$p$ valence band, and a large band gap to the Fe
$d$ conduction band. The lsd DOS shows a substantially reduced band
gap, which is between the occupied and unoccupied Fe impurity $d$
bands. So, while the SIC-LSD LS state is a charge transfer
insulator, the LSD gives rise to a Mott-Hubbard
insulator (see Fig. \ref{pic:res:DOSd}).
% Fe-3\emph{d} electrons form an isolated band above the Fermi level while the material remains an insulator.

\subsection{GGA-PAW calculations, Fe clustering \label{sec:res:vasp_clust}}

%\subsection{PAW-GGA methodology\label{sec:meth:gga}}
%
So far we have concentrated mostly on the correlated nature of Fe
$d$ electrons in describing electronic properties of doped MgO
within SIC-LSD approach.
%However, it is likely that the SIC-LSD
%results could be considerably influenced by a possible Fe impurity
%clustering.
Here we consider an importance of a possible Fe impurity clustering.
To study this, we have performed a number of
independent calculations for low Fe concentrations in
Mg$_{1-x}$Fe$_x$O, using the standard GGA
approximation\cite{perdew1996a} and the Vienna Ab-Initio Simulation
Package (VASP)\cite{kresse1996}. The latter provides the standard
projector augmented wave potentials for Mg, Fe and O with core
configurations 2p$^6$3s$^2$ (Mg), 3p$^6$3d$^7$4s$^1$ (Fe) and
2s$^2$2p$^4$ (O).\cite{kresse1999} In the actual calculations the
electronic optimization has been run with a self-consistency
threshold of 10$^{-9}$~eV, ionic optimization has been done with a
threshold of 10$^{-6}$~eV. The $4\times4\times4$-Monkhorst-Pack
scheme has been applied for the k-point sampling, producing four
irreducible k-points.\cite{monkhorst1976} The energy cut-off of the
plane wave basis set has been equal to 700~eV. In order to speed up
the evaluation of the non-local part of the potentials real space
projections have been used. A second order Methfessel-Paxton
smearing has been used ($\sigma$=0.04~eV), introducing an error in
electronic entropy, $T\, S_{\mathrm{el}}$, of at most
0.111~meV/atom.\cite{methfessel1989} The accuracy of the energy
differences between the HS and the LS state has been converged to
$1.3\times10^{-6}$~eV/atom. Note, however, that within the GGA
implementation all the electrons are treated on equal footing as
delocalized, with the HS state being described as a ferromagnetic
state, whilst the LS state as a non-magnetic state. Thus the two are
not directly comparable to the HS and LS states calculated within SIC-LSD,
as the concept of divalent and trivalent ions cannot be clearly defined
within GGA.
%Although GGA also introduces an unphysical SI, we do not correct for
%it in this work, as it is different than that of LSD, and not
%implemented within the standard VASP package.

To investigate the effects of Fe clusters, supercells containing 64
atoms have been set up, including one and four Fe impurities
($x_{\mathrm{Fe}}=3.125$ \%, $x_{\mathrm{Fe}}=$12.5 \%). In the
latter case, 30 distinct arrangements of Fe atoms
exist.\footnote{Note that this is a combinatorial problem. The
clusters can be characterized by the distance between the first Fe
impurity to the following ones. Five different distances can be
found in the 64 atomic cell. They can host 12, three, 12, three and
one Fe atom respectively. This gives one, five, 14, 30, 53 distinct
scenarios of one, two, three, four and five impurities,
respectively.} We have used the most extreme cases, where the Fe
atoms form a tetragonal cluster with an Fe-Fe distance of $\simeq$
2.69~\AA~ and where they avoid each other at a maximal distance of
$\simeq$ 5.37~\AA~ (at 100~GPa). The resulting three clusters (one
Fe atom, four Fe atoms at minimum distance, four Fe atoms at maximum
distance) have been explored performing spin-polarized and
non-spin-polarized GGA calculations (in the former case, a starting
magnetic moment of 4 $\mu_B$ has been assigned to each Fe atom). In
the spin-polarized case the Vosko-Wilk-Nusair interpolation for the
correlation part of the exchange-correlation functional has been
used,\cite{vosko1980}. % which enhanceses spin polarization.
The lowest energy structure has been found by comparing the enthalpies of
all the different structures.

The GGA-PAW calculations in the 64 atomic supercell find HS-LS
transition pressures of 26~GPa ($x=3.125$~\%), 30.3~GPa ($x=12.5$~\%,
diluted configuration of Fe), and 63.4~GPa ($x=12.5$~\%, clustered
configuration of Fe), as calculated from the intersection of the
enthalpy curves of the ferromagnetic and non-magnetic results.
\begin{figure*}[] %Plazierung
 \centering
 \epsfig{file=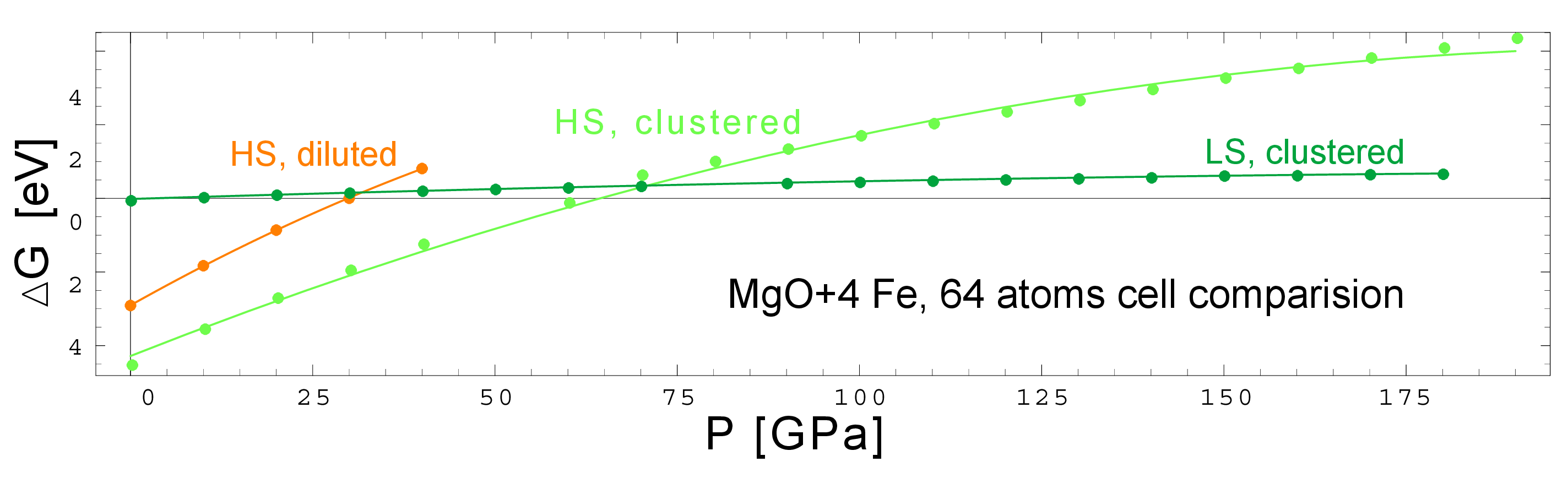, width=0.9\textwidth}  % scale=FACTOR, heigth=, width=, angle=
  \caption{ \label{pic:res:compare4_64}
  GGA-PAW calculations, 64 atomic supercell including four Fe atoms: Comparison of the enthalpies of
the different Fe and spin arrangements to the diluted LS setting. In
the clustered configuration Fe atoms are at the largest distance
possible. }
%\end{figure}
\end{figure*}
The lowest enthalpy structures always give a total magnetization of
either  $\mu=0.0\mu_B$ (non-magnetic) or $\mu=4.0\mu_B$
(ferromagnetic). Arrangements of Fe atoms with an intermediate spin
magnetic moment have been found in some cases (e.g. $\mu=3\mu_B$ in
the case of Fe clusters between 90~GPa and 185~GPa). However they
always turn out to be unstable because energetically lower lying
cluster configurations exist.

The transition pressure turns out to increase with the Fe
concentration. The two studied concentrations ($x=3.125\%$ and
$x=12.5\%$) allow to determine the linear dependence of the
transition pressure as

\begin{equation} \label{eq:vasp_clust:1}
P_{\mathrm{cr}}=24.6+0.45\,x\;.
\end{equation}
The relatively small difference to the experimental value of
$P_{\mathrm{cr}}= 28 + 0.63\, x$ can be explained, if it is assumed
that the statistical distribution of the Fe atoms in the
experimental sample will not be perfectly diluted, but in some cases
result in having Fe atoms at smaller distances (which dramatically
raises the critical pressure). However, this can also explain the
broadness of the experimentally found transition pressure: depending
on the local Fe configuration the spin magnetic moments collapse at
different pressures.

Figure \ref{pic:res:compare4_64} indicates that at low pressures
(where only the HS arrangments are of interest) the Fe atoms prefer
to form clusters, while at high pressures (where only the LS
arrangments are of interest) the Fe atoms prefer a diluted
configuration. Upon pressure release at the pressure of about 60~GPa
the LS diluted arrangement and the HS clustered arrangement
coincide. The different Fe arrangments in the high pressure and low
pressure regions might hamper this phase transition. Experimental
HS-LS transition pressures are predicted to depend on the pressure
the sample has been equilibrated at (the equilibration could be
enhanced through heat, which accelerates ionic diffusion). If the
equilibration is done at high pressure, upon pressure release the LS
configuration would be stable up to relatively low pressures
($\simeq$30~GPa). A sample equilibrated at low pressure on the other
hand would preserve its magnetization up to relatively high
pressures ($\simeq$60~GPa).

\section{Conclusions \label{sec:con}}

Our GGA results indicate a first order transition for all the
compositions studied, both for Mg$_{1-x}$Fe$_x$O and Fe$_{1-x}$O, in
good agreement with the notion in Ref.~\onlinecite{ono2007} for
Fe$_{1-x}$O. The electronic rearrangement of the Fe atoms leads to a
discontinuity of the first derivative of the enthalpy, i.e. the
volume (see Fig.~\ref{pic:res:compare4_64}). This affects various
physical properties of the HS and LS phases, such as the zero
pressure bulk modulus and the ground state volume.

In experiment the HS-LS transition appears to be smeared out over a
large pressure range,\cite{badro2003, speziale2005,lin2005} which in
the case of Fe$_{1-x}$O has led to the conjecture that the
transition might be second order.\cite{pasternak1997} We can not
support this unless this is a temperature induced smearing which our
$T=0$ calculations do not consider. The results for various Fe
arrangements, keeping the Fe concentration fixed, indicate that the
local arrangement of Fe strongly influences the spin transition
pressure of the Fe atoms. The statistical distribution of the Fe
atoms and the short range defect order can therefore smear out the
transition pressure over a large pressure range. As the HS-LS
transition is isosymmetric, according to Landau theory it can only
be first-order or (above the critical temperature) fully
continuous.\cite{christy1995, bruce1981}

We can confirm the experimentally found trend that the HS-LS
transition pressure increases with increasing Fe concentration. This
might be the result of Fe atoms which --- if situated closely enough
--- simultaneously optimize their electronic configurations. These
magnetic Fe clusters can not easily be demagnetized by pressure. The
increase of the Fe concentration leads to a statistical increase of
short Fe-Fe distances, which through the described mechanism could
in turn lead to an increase of the critical pressure.

It has been proposed that the HS-LS transition in
magnesiow\"{u}stite would lead to a discontinuity in the Earth's
lower mantle.\cite{badro2003} Although the transition appears to be
first order a discontinuity seems to be improbable: first the
sensitivity of the transition pressure to the (local) Fe
concentration could smear out the transition over a wide region in
depth. Second, temperature could further increase this effect and
shift the transition to even higher pressures, barring LS Fe$^{2+}$
from the Earth's interior.\cite{tsuchiya2006, sturhahn2005}

The transition pressures predicted by SIC-LSD appear to be higher
which might be a result of an
``overcorrection'' of the energy of the localized states. It has
been argued in Ref. \onlinecite{perdew2005} that a weighted
self-interaction correction would be desirable. The high transition
pressures calculated in this work might also be due to the fact that
the ASA does not allow relaxations of the atomic spheres, which
might be significant in the LS phase, where the ionic radius of Fe
is small.
At high pressures the LSD treatment can lead to low energy structures.
This could justify
the treatment of Mg$_{1-x}$Fe$_x$O at those pressures using standard
approximations for the exchange-correlation energy such as the GGA.
Apart from this it could also indicate a second phase transition
which involves a full delocalization of the Fe-3\emph{d} states.
Furthermore all the transitions considered in this paper
involve five or six localized electrons in the high volume phase and
no localized electrons in the low volume phase. Consideration of
valence fluctuations or
`intermediate' localization involving four, three, two and one
localized states could plausibly lead to a reduction of the
transition pressure.
The SIC-LSD formalism, implemented in the multiple scattering
theory, allows to consider static valence and spin fluctuations at
finite temperatures, by invoking CPA (coherent potential
approximation) and DLM (disordered local moments) approaches, which
was successfully applied to studying phase
diagrams.\cite{lueders2005, hughes2007} This would also be useful
for the further theoretical investigation of the present
geophysically relevant systems.
For the nearest future, it may be instructive to first apply SIC-LSD
in a full potential version in order to allow
for ionic relaxations to study the influence of distortions on
the ground state properties of these systems.

\section*{Acknowledgements \label{sec:ackno}}

The authors are indebted to A.R. Oganov who contributed considerably
to the concept of this work. DJA thanks A.R. Oganov and the ETH
Zurich Research fund for their support of this work (Grant No.
TH-27033). Supercomputers were provided by the Swiss National
Supercomputing Centre (CSCS) and ETH Zürich. WMT and ZS would like
to acknowledge useful discussions with Drs. Axel Svane and Leon
Petit.

%
%\appendix
%\section{ \label{sec: }}

%\bibliography{/cea/home/pmc/adams/biblio/bibliography}
\bibliography{bibn}

\end{document}